\documentclass[letterpaper]{article} 
\usepackage{aaai2026}
\usepackage{times}  
\usepackage{helvet}  
\usepackage{courier}  
\usepackage[hyphens]{url}  
\usepackage{graphicx} 
\usepackage{natbib}  
\usepackage{caption} 
\usepackage{algorithm}
\usepackage{algorithmic}

\usepackage{newfloat}
\usepackage{listings}
\usepackage{booktabs}
\usepackage{multirow}
\usepackage{tcolorbox}
\usepackage{amsmath}
\usepackage{amssymb}
\usepackage{makecell}
\DeclareCaptionStyle{ruled}{labelfont=normalfont,labelsep=colon,strut=off} 
\floatstyle{ruled}
\newfloat{listing}{tb}{lst}{}
\floatname{listing}{Listing}
\title{E\textsuperscript{3}-Rewrite: Learning to Rewrite SQL for Executability, Equivalence,\\and Efficiency}

\author {
    Dongjie Xu\textsuperscript{\rm 1}\thanks{Email: djxu@stu.suda.edu.cn},
    Yue Cui\textsuperscript{\rm 2},
    Weijie Shi\textsuperscript{\rm 2},
    Qingzhi Ma\textsuperscript{\rm 1},
    Hanghui Guo\textsuperscript{\rm 3},
    Jiaming Li\textsuperscript{\rm 4},
    Yao Zhao\textsuperscript{\rm 5},
    Ruiyuan Zhang\textsuperscript{\rm 2},
    Shimin Di\textsuperscript{\rm 6},
    Jia Zhu\textsuperscript{\rm 3},
    Kai Zheng\textsuperscript{\rm 7},
    Jiajie Xu\textsuperscript{\rm 1}
}
\affiliations {
    \textsuperscript{\rm 1}Soochow University \quad
    \textsuperscript{\rm 2}Hong Kong University of Science and Technology \quad
    \textsuperscript{\rm 3}Zhejiang Normal University \quad
    \textsuperscript{\rm 4}ByteDance Inc. \quad
    \textsuperscript{\rm 5}Alibaba Group \quad
    \textsuperscript{\rm 6}Southeast University \quad
    \textsuperscript{\rm 7}University of Electronic Science and Technology of China
}

\usepackage{bibentry, enumitem}

\begin{document}

\maketitle

\begin{abstract}
SQL query rewriting aims to reformulate a query into a more efficient form while preserving equivalence. Most existing methods rely on predefined rewrite rules. However, such rule-based approaches face fundamental limitations: (1) fixed rule sets generalize poorly to novel query patterns and struggle with complex queries; (2) a wide range of effective rewriting strategies cannot be fully captured by declarative rules. To overcome these issues, we propose using large language models (LLMs) to generate rewrites. LLMs can capture complex strategies, such as evaluation reordering and CTE rewriting. Despite this potential, directly applying LLMs often results in performance regressions or non-equivalent rewrites due to a lack of execution awareness and semantic grounding. To address these challenges, We present \textbf{E\textsuperscript{3}-Rewrite}, an LLM-based SQL rewriting framework that produces executable, equivalent, and efficient queries. It integrates two core components: a context construction module and a reinforcement learning framework. First, the context module leverages execution plans and retrieved demonstrations to build bottleneck-aware prompts that guide inference-time rewriting. Second, we design a reward function targeting executability, equivalence, and efficiency, evaluated via syntax checks, equivalence verification, and cost estimation. Third, to ensure stable multi-objective learning, we adopt a staged curriculum that first emphasizes executability and equivalence, then gradually incorporates efficiency. Across multiple SQL benchmarks, our experiments demonstrate that E\textsuperscript{3}-Rewrite can shorten query execution time by as much as 25.6\% relative to leading baselines, while also producing up to 24.4\% more rewrites that meet strict equivalence criteria. These gains extend to challenging query patterns that prior approaches could not effectively optimize.
\end{abstract}


\section{Introduction}

Efficient query processing is a central goal of modern database systems, and SQL query rewriting plays a key role in improving performance~\cite{DBLP:journals/pvldb/Li19,DBLP:journals/tkde/ZhouCLS22}. The goal is to rewrite a given SQL query into an equivalent but more efficient form~\cite{DBLP:conf/sigmod/WangZYDHDT0022}. To be practically useful, a rewritten query must satisfy three criteria: (1) \textbf{Executability}: the rewritten query must conform to SQL syntax and execute correctly within a target DBMS; (2) \textbf{Equivalence}: the rewritten query must preserve the original query's semantics and produce identical results; (3) \textbf{Efficiency}: the rewritten query should reduce execution time, and the overhead of the rewriting process itself should be justified by the resulting performance gains. Only when all three conditions are met can a rewrite be deemed both theoretically valid and practically beneficial.

Most SQL rewriting methods follow a rule-based paradigm: they apply predefined rewrite rules to convert an input query into an equivalent but more efficient form~\cite{DBLP:journals/pvldb/BaiA023,DBLP:conf/sigmod/BegoliCHML18}. Early systems rely on manually crafted rules and heuristics, which generalize poorly to diverse query patterns. To improve flexibility, later methods frame rewriting as a rule selection task. For example, LearnedRewrite~\cite{DBLP:journals/pvldb/ZhouLCF21} uses Monte Carlo Tree Search guided by cost estimators to explore rule combinations. Recently, large language model (LLM)-based systems such as LLM-R²~\cite{DBLP:journals/pvldb/LiYWCB24} and R-Bot~\cite{DBLP:journals/corr/abs-2412-01661} leverage LLM to model queries and suggest appropriate rules via prompting or retrieval. While these approaches reduce manual effort, they remain constrained by the fixed coverage of predefined rule sets. 

More fundamentally, rule-based rewriting systems suffer from limitations that restrict their expressiveness, composability, and adaptability. First, fixed rule sets lack expressiveness and struggle to capture structural rewrites beyond predefined patterns~\cite{DBLP:conf/sigmod/BegoliCHML18,DBLP:journals/corr/abs-2403-09060}. Second, rule dependencies form an entangled and fragile search space, posing challenges for both traditional search-based methods~\cite{DBLP:journals/pvldb/ZhouLCF21} and LLM-driven prompting~\cite{DBLP:journals/pvldb/LiYWCB24}, often resulting in brittle recommendations. Third, effective rewriting strategies inherently fall outside the expressive capacity of declarative rules~\cite{DBLP:journals/pvldb/DongLZYMW23,DBLP:journals/pvldb/LeisGMBK015}, especially those involving Common Table Expressions (CTEs) or requiring fine-grained control over evaluation order and sub-plan reuse~\cite{DBLP:conf/sigmod/WangZYDHDT0022}. Most importantly, existing methods cannot adapt to new query structures or workloads over time, as they lack mechanisms to learn from execution feedback and refine their rewriting strategies. A recent concurrent approach, QUITE~\cite{DBLP:journals/corr/abs-2506-07675}, explores a training-free, feedback-aware multi-agent framework that iteratively refines rewrites at inference time via a hybrid SQL corrector and hint injection. While effective in certain settings, such inference-time correction strategies may require multiple iterations and lack integrated training-time feedback, making it challenging to jointly optimize executability, equivalence, and efficiency.

To overcome the limitations of rule-based methods, we propose a fundamentally different approach: training an LLM to directly generate equivalent and efficient SQL rewrites without relying on explicit rules. Unlike standard supervised training tasks of LLMs, SQL rewriting lacks labeled targets, as optimal rewrites must be identified through execution. The quality of a rewrite depends on executability, equivalence, and runtime efficiency, properties that are non-differentiable, delayed, and often conflicting. These characteristics make supervised learning ineffective. We therefore adopt reinforcement learning (RL), which optimizes rewriting behavior through execution feedback, using rewards from parsing results, equivalence checks, and cost measurements. Achieving this goal introduces three key challenges: (1) Due to the lack of execution awareness and semantic grounding, LLMs struggle to generate SQL rewrites that are executable, equivalent, and performance-optimized; (2) The learning process lacks stable and interpretable signals that connect rewriting behaviors with both semantic equivalence and execution performance gain; and (3) Simultaneously optimizing equivalence and efficiency from the beginning leads to unstable learning, as the reward signals are often conflicting and hard to balance.

To address the above challenges, we introduce \textbf{E\textsuperscript{3}-Rewrite}, an LLM-based SQL rewriting framework optimized for executability, equivalence and efficiency. To tackle the first challenge, we incorporate query execution plans into the input context to reveal logical structure and performance bottlenecks. To further enhance structural awareness and generalization, we retrieve analogous rewrites from a hybrid structure-semantic demonstration library during inference. To address the second challenge, where rewriting quality cannot be directly supervised, we design an RL framework that optimizes the model with execution-driven rewards. These rewards jointly reflect executability, equivalence, and efficiency, enabling the model to improve through feedback rather than fixed labels. For the third challenge, which involves instability from multiobjective optimization, we adopt a curriculum-based training strategy that first focuses on generating executable and equivalent rewrites, and later introduces latency as an optimization signal. This staged learning helps mitigate reward interference and supports stable policy updates. 

\noindent Our main contributions are:
\begin{itemize}[leftmargin=1.5em, label=\raisebox{-0.1ex}{\fontsize{12pt}{12pt}\selectfont$\bullet$}]
  \item We propose E\textsuperscript{3}-Rewrite, the first LLM-based SQL rewriting framework that generates executable, equivalent, and efficient queries via end-to-end optimization without relying on rule sets.
  
  \item We develop a RL framework with a curriculum-based multi-stage training strategy to enable stable optimization across executability, equivalence, and efficiency.

  \item We introduce execution-guided context construction and hybrid retrieval based on query plans and structure-semantic similarity. These modules expose logical structures, highlight performance bottlenecks, and support effective demonstration reuse for better generalization.

  \item We conduct extensive experiments on widely used SQL benchmarks to demonstrate that E\textsuperscript{3}-Rewrite significantly outperforms state-of-the-art query rewriting methods in both execution efficiency and equivalence preservation.

\end{itemize}

\section{Related Work}

\noindent \textbf{Query Rewriting.}  
Existing query rewriting methods have evolved from rule-based systems to learning-based and LLM-augmented approaches.\textit{(1) Heuristic Rule-based Systems.}  Early systems like PostgreSQL~\cite{postgresql2025} and Volcano~\cite{DBLP:conf/icde/GraefeM93} rely on fixed or heuristic rule sequences. These approaches often neglect inter-rule dependencies, leading to blind or inefficient search.
\textit{(2) Learning-based Rule Selection.}  
LearnedRewrite~\cite{DBLP:journals/pvldb/ZhouLCF21} treats rewriting as a rule search problem and uses MCTS with learned cost models. However, generalization across schemas remains limited due to reliance on schema-specific reward estimators.
\textit{(3) LLM-Augmented Rewriting.} LLM-R²~\cite{DBLP:journals/pvldb/LiYWCB24} and R-Bot~\cite{DBLP:journals/corr/abs-2412-01661} leverage LLMs for rule retrieval or recommendation via prompting. QUITE~\cite{DBLP:journals/corr/abs-2506-07675} adopts a training-free, feedback-aware multi-agent pipeline that combines a structured knowledge base, hybrid SQL corrector, and hint injection to iteratively refine rewrites. While promising, such inference-time correction frameworks often struggle to simultaneously ensure semantic equivalence and achieve performance gains, and may require multiple iterative inference rounds to converge on a satisfactory rewrite, which significantly increases execution time. These limitations motivate the need for a fully learnable, execution-driven rewriting framework that generalizes beyond rule enumeration.

\noindent \textbf{LLM-Based SQL Generation.}  
Parallel to rewriting, LLMs have achieved remarkable performance in NL2SQL tasks, including DIN-SQL~\cite{DBLP:conf/nips/PourrezaR23}, MAC-SQL~\cite{DBLP:conf/coling/WangR0LBCYZYSL25}, and COGSQL~\cite{DBLP:conf/aaai/YuanTCSCL25}. These systems adopt prompt engineering, chain-of-thought reasoning, or multi-agent collaboration to enhance generation quality.However, most focus solely on producing executable queries, and evaluate correctness via execution consistency~\cite{DBLP:conf/emnlp/YuZYYWLMLYRZR18}. They do not ensure strict semantic equivalence with a reference query—an essential aspect in SQL rewriting where input-output fidelity is critical.

\begin{figure*}[t]
    \centering
    \includegraphics[width=\linewidth]{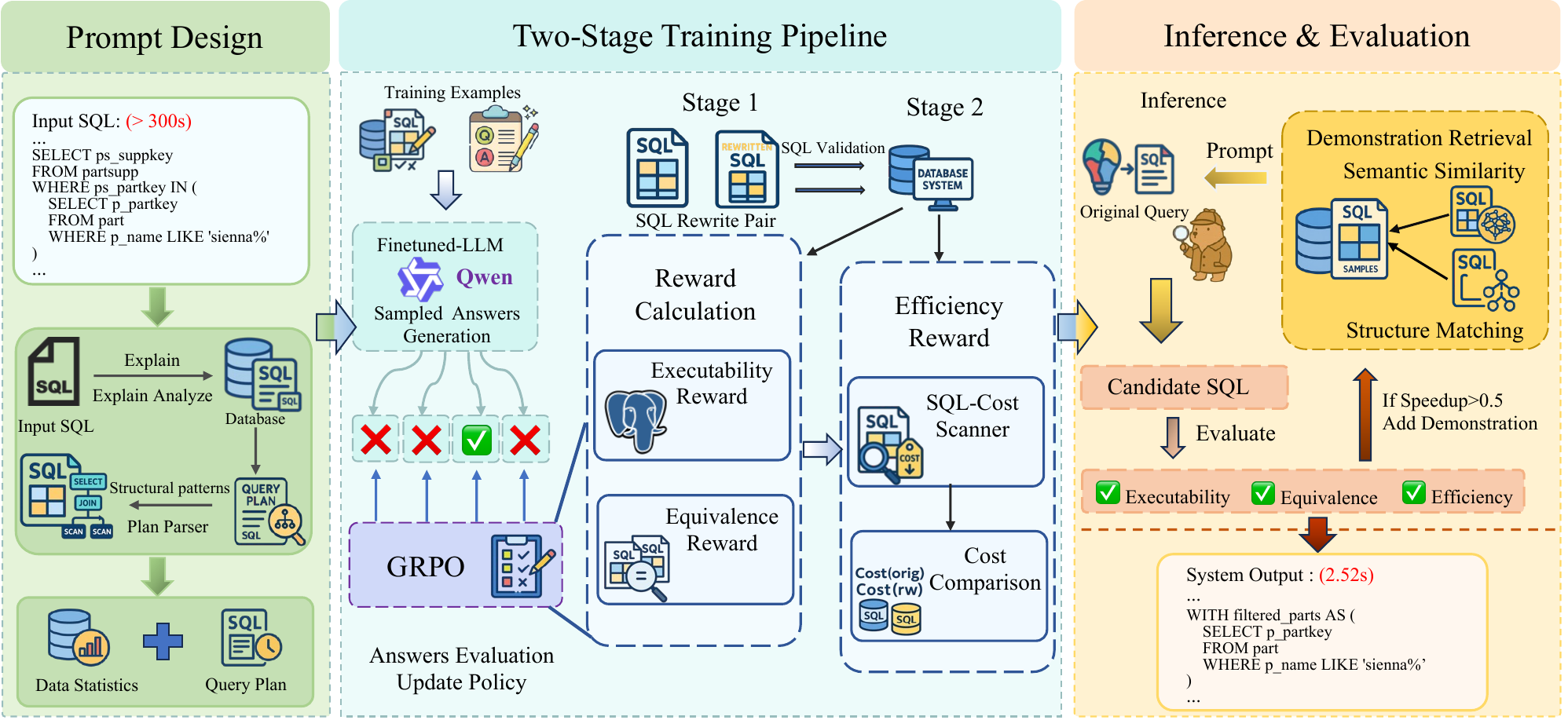}
    \caption{Overview of our E³-Rewrite. The system introduces execution hints into the prompt using parsed query plans, then optimizes a finetuned LLM with two-stage GRPO training that incorporates executability, equivalence, and efficiency rewards. During inference, E³-Rewrite retrieves hybrid demonstrations based on structure and semantics to guide rewriting, and updates its memory pool with rewrites that yield substantial execution gains.}
    \label{fig:overview}
\end{figure*}

\section{Problem Formulation}

We aim to develop an SQL rewriting system that rewrites an input query into an equivalent yet more efficient form. Formally, given an input query $q$, the goal is to generate a rewritten query $q'$ that satisfies the following conditions:

\begin{itemize}
  \item \textbf{Executability}: $q'$ conforms to SQL syntax and executes successfully on the target DBMS.

  \item \textbf{Equivalence}: $q \equiv q'$, i.e., for any database instance $D$, the result sets $q(D)$ and $q'(D)$ are identical, ensuring query equivalence.

  \item \textbf{Efficiency}: Under the same execution environment, the runtime of $q'$ is lower than that of $q$, i.e., $T(q') < T(q)$.
\end{itemize}

This formulation sets the foundation for learning-based approaches to SQL rewriting, where the objective is to automatically generate rewrites that meet the above criteria.

\section{Method}

\subsection*{The Overview of E\textsuperscript{3}-Rewrite}
E\textsuperscript{3}-Rewrite propose an end-to-end SQL rewriting framework that combines large language models with RL and demonstration reuse to generate executable, equivalent, and efficient SQL rewrites. An overview of the full workflow is shown in Figure~\ref{fig:overview}.

The system consists of three major components. In the first stage, \textbf{Prompt Design}, we analyze the input SQL using EXPLAIN or EXPLAIN ANALYZE to extract its query execution plan. The plan is then linearized into an indented text format that preserves the operator hierarchy and runtime bottlenecks. This execution hint is prepended to the original query, allowing the model to understand structural inefficiencies such as full scans or unindexed joins.

In the second stage, \textbf{Two-Stage Reinforcement Learning}, the language model is fine-tuned to generate improved rewrites guided by reward signals. For each input, the model samples multiple candidates which are scored based on executability, equivalence, and estimated execution cost. We apply the GRPO algorithm to update the model using relative advantages across candidate groups. To encourage stable training, we adopt a curriculum strategy: the first stage emphasizes correctness through executability and equivalence rewards, and the second stage incorporates cost-based feedback to optimize efficiency.

The third stage, \textbf{Inference \& Evaluation}, deploys the trained model for inference-time rewriting. To provide guidance, a hybrid demonstration retriever selects past successful rewrites based on structural and semantic similarity. If a newly generated query achieves semantic equivalence and improves execution cost beyond a given threshold, it is added back into the demonstration pool for future reuse.

\subsection*{Execution Hint Injection}

To incorporate structural and execution-level signals into model input, we treat the query execution plan as an \textit{execution hint} that reflects how the DBMS interprets and optimizes a given SQL query. Compared to raw SQL statements, execution plans expose operator choices, join orders, index usage, and runtime bottlenecks, making them valuable guides for identifying optimization opportunities.

We extract execution plans using standard EXPLAIN and EXPLAIN ANALYZE~\cite{postgresql2025}. EXPLAIN provides a static plan with cost estimates from the optimizer, while EXPLAIN ANALYZE executes the query and yields runtime-level feedback including operator latencies and row counts. During training, we use EXPLAIN ANALYZE to capture realistic performance traces; during inference, we switch to EXPLAIN to reduce latency while preserving structural patterns.

To make these hints model-consumable, we linearize the hierarchical plan into a flattened, indented text format that preserves operator nesting and structural flow. The resulting hint is prepended to the SQL query, forming a plan-aware input context. This injection enables the model to identify inefficiencies such as full scans, non-indexed joins, or redundant filtering, and to generate more efficient rewrites guided by execution-level signals.

\begin{figure}[t]
    \centering
    \includegraphics[width=\linewidth]{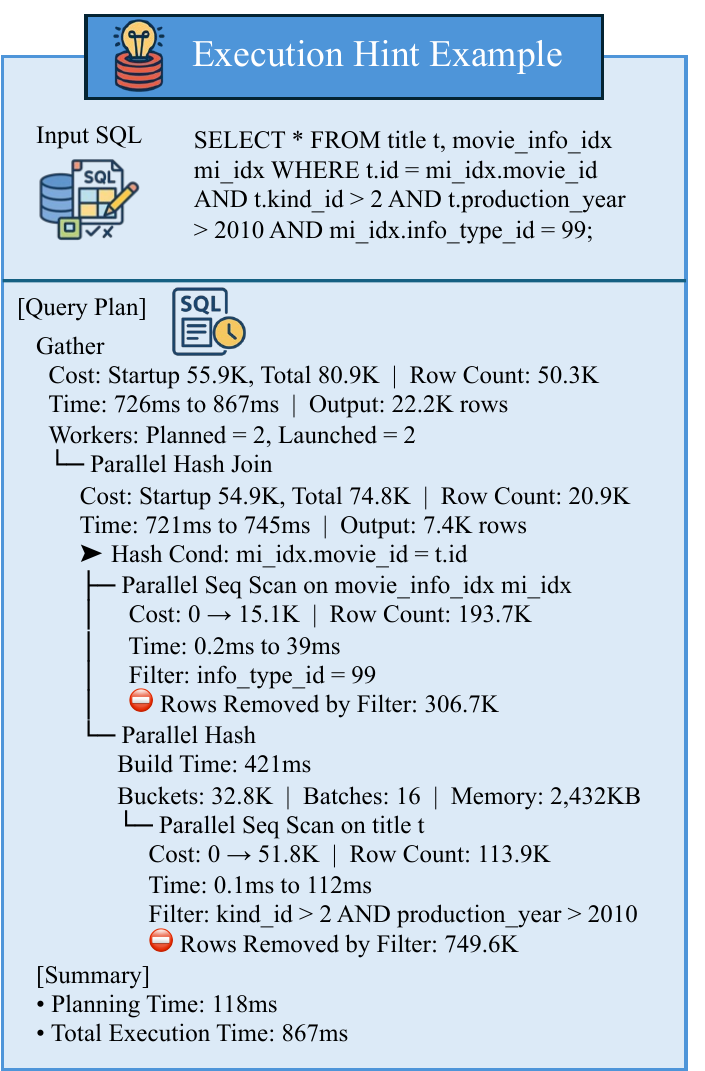}
    \caption{Execution hint example.}
    \label{fig:hint-example}
\end{figure}

As shown in Figure~\ref{fig:hint-example}, the execution hint reveals a multi-level plan with explicit bottlenecks, such as a large number of rows removed by post-filtering and unindexed sequential scans, thereby providing the model with actionable structural cues for rewriting.

\subsection*{Reinforcement Learning for SQL Rewriting}
To enable self-improving SQL rewriting, we fine-tune foundation LLMs with RL guided by execution-aware reward signals. Traditional methods like PPO~\cite{DBLP:journals/corr/SchulmanWDRK17} struggle with discrete, sparse, and conflicting rewards, which are common in SQL rewriting. We therefore adopt the Group-Relative Policy Optimization (GRPO) algorithm~\cite{DBLP:journals/corr/abs-2501-12948}, which eliminates the need for a value network and stabilizes learning via intra-group comparisons of sampled rewrites. This design makes GRPO well-suited for our setting, where rewrite quality depends on non-differentiable and delayed execution feedback.

\noindent \textbf{GRPO Overview.} For each input query $q$, the model samples $N$ candidate rewrites $\{q'_1, q'_2, \ldots, q'_N\}$. Each candidate is evaluated with a scalar reward $r_i$, and its relative advantage, denoted as $A_i$, is computed by group-wise normalization.

\begin{equation}
A_i = \frac{r_i - \text{mean}({r_1, \ldots, r_N})}{\text{std}({r_1, \ldots, r_N})}
\end{equation}

GRPO minimizes a clipped loss by weighting token log-likelihoods with normalized relative advantages.

{\small
\begin{equation}
\begin{split}
\mathcal{L}_{\text{GRPO}}(\theta) =
- \frac{1}{\sum_{i=1}^{G} |o_i|} 
\sum_{i=1}^{G} \sum_{t=1}^{|o_i|}
\left[
\min\left(
\frac{\pi_\theta(o_{i,t} \mid q, o_i, <t)}
     {\pi_{\text{old}}(o_{i,t} \mid q, o_i, <t)}
\right),\right. \\
\left.
\text{clip}\left(
\frac{\pi_\theta(o_{i,t} \mid q, o_i, <t)}
     {\pi_{\text{old}}(o_{i,t} \mid q, o_i, <t)},
1 - \epsilon,\ 1 + \epsilon
\right) \hat{A}_{i,t}
- \beta\, \mathbb{D}_{\text{KL}} \left[ \pi_\theta \| \pi_{\text{ref}} \right]
\right]
\end{split}
\end{equation}
}

where $\pi_\theta$ is the current policy, $\pi_{\text{old}}$ is the policy before update, and $\pi_{\text{ref}}$ is a reference policy for KL regularization. The hyperparameter $\beta$ controls the regularization strength. The scalar $\epsilon$ is a clipping threshold that limits the ratio between new and old policies, and stabilizes the optimization. 

\noindent \textbf{Execution-Aware Reward Function.} To guide learning with meaningful execution feedback, we define a composite reward that combines three objectives:
\begin{equation}
R_{\text{total}} = \lambda_{\text{eq}} R_{\text{eq}} + \lambda_{\text{exec}} R_{\text{exec}} + \lambda_{\text{perf}} R_{\text{perf}}
\end{equation}

$\bullet$ \textbf{Executability} ($R_{\text{exec}}$): We use EXPLAIN to pre-execute the query in the target DBMS. If the query passes both parsing and semantic analysis, we set $R_{\text{exec}} = 1$; otherwise, $R_{\text{exec}} = 0$.

$\bullet$ \textbf{Equivalence} ($R_{\text{eq}}$): We adopt a three-stage verification pipeline to determine whether the rewritten query is equivalent to the original query. First, we employ \textsc{QED-Solver}~\cite{DBLP:journals/pvldb/WangPC24}, a formal SQL equivalence checker, to determine equivalence. If the result is positive, we set $R_{\text{eq}} = 1$; if explicitly negative, we set $R_{\text{eq}} = 0$. If the solver returns an unknown verdict (e.g., timeout or unsupported constructs), we invoke a lightweight LLM-based semantic judgment to assess equivalence. For cases where the LLM output is inconclusive, we fall back to executing both queries and compare their outputs over sampled database instances. Only when the execution results match exactly, we set $R_{\text{eq}} = 1$; else, we set $R_{\text{eq}} = 0$.

$\bullet$ \textbf{Performance Improvement} ($R_{\text{perf}}$): We extract the estimated execution cost from EXPLAIN and define:
  \begin{equation}
  R_{\text{perf}} = \max \left(0, \frac{T(q) - T(q')}{T(q)} \right)
  \end{equation}
  where $T(\cdot)$ denotes the optimizer-estimated cost. This reward encourages performance improvement without compromising executability or equivalence.

\noindent \textbf{Curriculum-Guided Training.}
To stabilize learning and mitigate reward conflicts between executability, equivalence, and efficiency, we adopt a curriculum R L strategy~\cite{DBLP:conf/icml/BengioLCW09,DBLP:journals/corr/abs-2504-13592,DBLP:journals/corr/abs-2503-12759}. The training process is split into two stages:

$\bullet$ \textbf{Stage 1: Correctness-first Learning.} We activate only $R_{\text{eq}}$ and $R_{\text{exec}}$, guiding the model to produce SQL rewrites that are executable and equivalent to the original query.

$\bullet$ \textbf{Stage 2: Optimization-aware Refinement.} Once the model consistently satisfies correctness constraints, we enable $R_{\text{perf}}$ to encourage performance-oriented rewrites. Training proceeds with the full reward function, jointly optimizing for equivalence and efficiency.

$\bullet$ \textbf{(Optional) Rehearsal.} To prevent degradation in equivalence or executability, a small proportion of Stage 1 examples are periodically revisited during Stage 2.

This progressive reward schedule allows the model to first acquire a robust understanding of SQL executability and equivalence, before shifting attention to optimization focused on reducing execution latency. Combined with GRPO’s group level feedback mechanism, this design yields a stable rewriting policy informed by execution behavior, capable of generalizing to diverse and complex SQL workloads.

\subsection*{Hybrid Demonstration Retrieval}

To enable effective demonstration-based rewriting, we construct a retrieval module that selects the most relevant SQL rewrite examples from a maintained demonstration pool. The similarity between a new input query and existing examples is computed using a hybrid metric that combines both structural and semantic perspectives.

Let $q$ denote the input SQL query and $\mathcal{C} = \{q_i\}_{i=1}^{N}$ be the demonstration pool. For each $q_i \in \mathcal{C}$, we compute the overall similarity $\text{Sim}(q, q_i)$ as:
\begin{equation}
\text{Sim}(q, q_i) = \alpha \cdot \text{Sim}_{\text{struct}}(q, q_i) + (1 - \alpha) \cdot \text{Sim}_{\text{sem}}(q, q_i)
\end{equation}
where $\alpha \in [0, 1]$ is a balancing coefficient between the structural and semantic components.

\noindent \textbf{Structural Similarity.}
Following prior work on tree-based evaluation for SQL queries~\cite{DBLP:conf/icse/SongETLKBBBG24}, we define structural similarity based on the abstract syntax tree (AST) edit distance. Each SQL query is parsed into an operator-level AST $T(q)$, and the similarity is computed as:
\begin{equation}
\text{Sim}_{\text{struct}}(q, q_i) = 1 - \frac{\text{TED}(T(q), T(q_i))}{\max(|T(q)|, |T(q_i)|)}
\end{equation}
Here, $\text{TED}(\cdot, \cdot)$ denotes the tree edit distance computed via the APTED algorithm, and $|T(\cdot)|$ is the number of AST nodes. The score is normalized to $[0, 1]$, with higher values indicating stronger structural resemblance.

\noindent \textbf{Semantic Similarity.}
To measure the semantic alignment between queries, we compute the cosine similarity of their contextual embeddings:
\begin{equation}
\text{Sim}_{\text{sem}}(q, q_i) = \cos\left( \mathbf{e}_q, \mathbf{e}_{q_i} \right)
\end{equation}
where $\mathbf{e}_q, \mathbf{e}_{q_i} \in \mathbb{R}^d$ are dense representations derived from a pretrained transformer encoder~\cite{DBLP:conf/emnlp/ReimersG19}. This component captures query-level intent beyond syntactic forms.

\noindent \textbf{Demonstration Retrieval and Pool Update.}
For a given input query $q$, we retrieve the top-$k$ examples from $\mathcal{C}$ with the highest similarity scores as in-context demonstrations. The demonstration pool is dynamically updated: whenever a new rewrite yields equivalent and achieves a significant performance gain (e.g., $>1.5\times$ speedup), the original and rewrite pair is added to $\mathcal{C}$. This continual update mechanism ensures that the retrieval base evolves with high quality, generalizable rewrites.

\section{Experiment}
\begin{table*}[t]
\centering
\setlength{\tabcolsep}{4pt}
\begin{tabular}{l|ccc|ccc|ccc}
\toprule
\multirow{2}{*}{Query Latency (s)} & \multicolumn{3}{c|}{TPC-H} & \multicolumn{3}{c|}{IMDB} & \multicolumn{3}{c}{DSB} \\
\cmidrule(lr){2-4} \cmidrule(lr){5-7} \cmidrule(lr){8-10}
 & Average & Median & p90 & Average & Median & p90 & Average & Median & p90 \\
\midrule
\textbf{Origin} & 78.81 & 10.75 & 300.00 & 5.74 & 1.54 & 18.53 & 38.83 & 3.53 & 300.00 \\
\midrule
\textbf{LearnedRewrite} & 41.34 & 9.73 & 103.41 & 5.73 & 1.54 & 18.51 & 32.61 & 3.19 & 98.67 \\
\textbf{LLM-R2} & 54.76 & 10.02 & 300 & 5.32 & 1.47 & 17.44 & 20.12 & 2.96 & 48.67 \\
\textbf{R-Bot} & 39.89 & 9.15 & 84.27 & 5.54 & 1.52 & 18.07 & 27.75 & 3.12 & 89.96 \\
\textbf{LLM only (GPT-4o)} & 74.92 & 10.72 & 300.00 & 5.72 & 1.54 & 18.53 & 38.44 & 3.53 & 300.00 \\
\midrule
\textbf{E\textsuperscript{3}-Rewrite (Qwen)} & \textbf{29.67} & \textbf{8.63} & \textbf{51.37} & \textbf{5.01} & \textbf{1.44} & \textbf{16.71} & \textbf{16.93} & \textbf{2.46} & \textbf{38.47} \\
\textbf{E\textsuperscript{3}-Rewrite (LLaMA)} & 33.72 & 8.91 & 66.94 & 5.19 & 1.46 & 17.13 & 18.06 & 2.57 & 46.33 \\
\bottomrule
\end{tabular}
\caption{Different Methods on Query Latency (s)}
\label{tab:main-latency}
\end{table*}

\subsection*{Experiment Setting}
\textbf{Dataset.} We evaluate our system on three widely used SQL rewriting benchmarks: 
(1) \textbf{TPC-H}~\cite{tpch-toolkit}, a standard decision support benchmark consisting of 22 parameterized query templates. We generate approximately 2,000 queries over a 10GB database using the official toolkit. 
(2) \textbf{IMDB (JOB)}~\cite{DBLP:conf/acl/MaasDPHNP11,DBLP:journals/pvldb/LeisGMBK015}, a benchmark derived from the Join Order Benchmark, containing 2,000 analytical queries over the IMDB schema with diverse join patterns and predicates. 
(3) \textbf{DSB}~\cite{DBLP:journals/pvldb/DingCGN21}, which extends TPC-DS to model modern decision support scenarios. It includes 2,000 complex queries featuring nested structures, aggregations, and patterns that are particularly challenging for query optimizers.

\noindent \textbf{LLM.} We experiment with three representative large language models. For our method, we fine-tune open-source models from the Qwen3 and LLaMA4 families. We also include GPT-4o, a commercial closed-source model that serves as the backbone for several LLM-based baselines, including LLM-only, LLM-R\textsuperscript{2}, and R-Bot.

\noindent \textbf{Baselines.} We compare our method with the following representative SQL rewriting systems:  
(1) \textbf{Learned Rewrite (LR)}~\cite{DBLP:journals/pvldb/ZhouLCF21}, which uses MCTS guided by a learned cost model to search for the best rule sequences;  
(2) \textbf{LLM-R²}~\cite{DBLP:journals/pvldb/LiYWCB24}, which prompts GPT-3.5 with demonstrations and applies rewrite rules via Apache Calcite;  
(3) \textbf{R-Bot}~\cite{DBLP:journals/corr/abs-2412-01661}, a retrieval-augmented LLM system that performs step-by-step rule selection with reflective reasoning;  
(4) \textbf{LLM only}~\cite{li2023can}, which directly generates rewritten queries from instructions and schema. If the output is invalid, we retain the original query for fairness.

\noindent \textbf{Evaluation Metrics.} We evaluate the quality of rewritten queries using the following metrics:  (1) \textbf{Query Latency}, which measures the execution time of the rewritten query. (2) \textbf{Equivalence Rate}, defined as the proportion of rewritten queries that produce identical output tuples as the original queries when executed on the same database instance. (3) \textbf{Improved Queries}, which counts the number of rewrites that yield a significant performance gain, defined as at least a 10\% reduction in execution time relative to the original query. 

For each query, we conduct five executions and calculate the average after excluding the highest and lowest values. For each metric, we report the average, median, and 90th percentile (p90). To ensure fairness, we impose a maximum execution timeout of 300 seconds. Queries that exceed this threshold are assigned a latency of 300 seconds.

\noindent \textbf{System Environment.} All queries are executed on PostgreSQL v14. All experiments are conducted on a server equipped with 515 GB RAM, a 3.0GHz 24-core Intel Xeon CPU, and 8 NVIDIA A100 GPUs (80GB each). Detailed training configurations and hyperparameter settings are provided in the supplementary material.

\subsection*{Execution Efficiency Evaluation}
\begin{figure}[t]
    \centering
    \includegraphics[width=\linewidth]{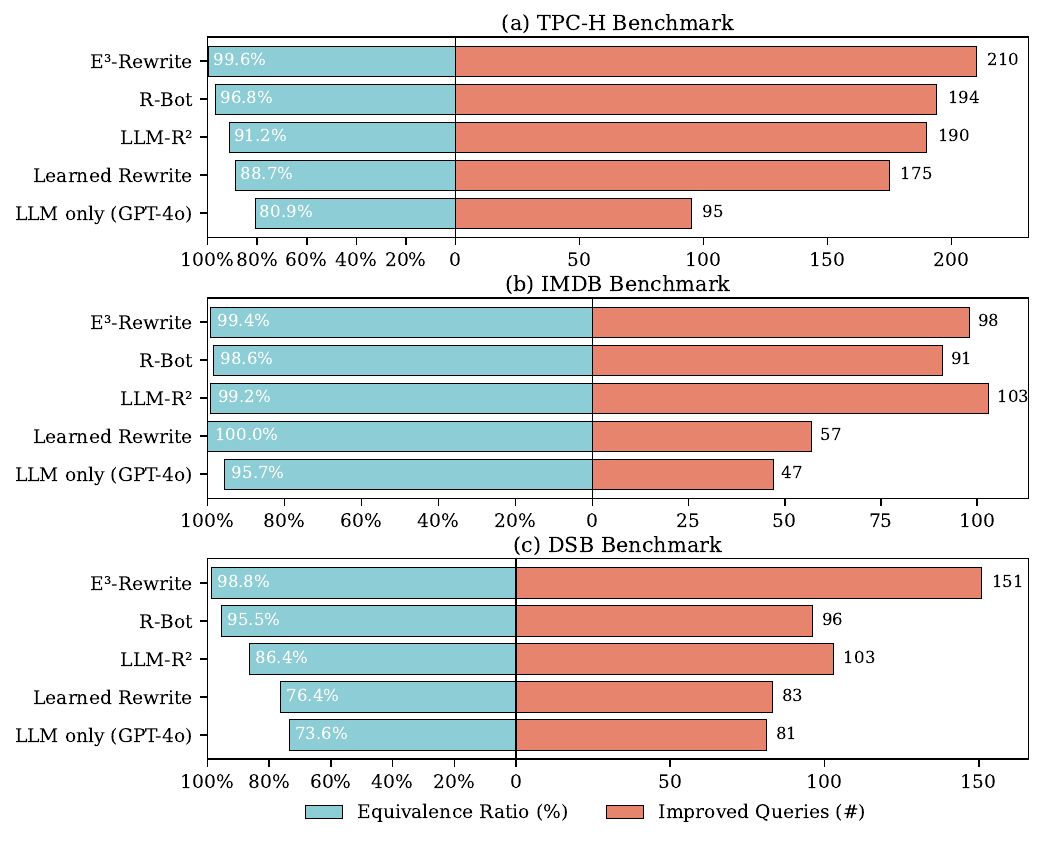}
    \caption{Equivalence Ratio and Improvement Counts of SQL Rewrites across Different Benchmarks.}
    \label{fig:rewrite_equivalence_improvement}
\end{figure}
We begin by evaluating the execution efficiency of different rewriting methods. Query latency is used as the primary metric, reflecting the execution cost of a rewritten SQL query. Each query is executed five times, and the average latency is computed after discarding the highest and lowest values. Table~\ref{tab:main-latency} summarizes the results on three benchmarks: TPC-H, IMDB, and DSB.

E\textsuperscript{3}-Rewrite (Qwen) achieves the lowest latency across all datasets. On TPC-H, it reduces the average latency from 78.81s (Original) to 29.67s and the p90 latency from 300.00s to 51.37s. While LearnedRewrite shows moderate improvement, it is limited by fixed rule coverage. LLM-R\textsuperscript{2} improves average latency but fails to reduce high-latency outliers. R-Bot performs well on IMDB but struggles on DSB. In contrast, E\textsuperscript{3}-Rewrite consistently reduces both average and p90 latency, showing strong robustness under complex and costly queries.

These gains are driven by three key components of our framework. First, execution hints reveal runtime inefficiencies (e.g., full scans, nested loops), guiding the model toward more optimized rewrites. Second, GRPO-based RL encourages rewrites that are both correct and performant. Third, our hybrid demonstration retrieval improves generalization by providing structurally and semantically relevant examples.

In summary, E\textsuperscript{3}-Rewrite significantly improves execution efficiency over prior methods, particularly on complex decision-support workloads.

\subsection*{Rewriting Quality and Model Behavior Analysis}

\begin{table}[t]
\centering
\renewcommand{\arraystretch}{1.2}
\begin{tabular}{lccc}
\toprule
\makecell{Method} & 
\makecell{Average\\Latency (s)} & 
\makecell{Improved\\Queries} & 
\makecell{Equivalence\\Ratio} \\
\midrule
\makecell[l]{E\textsuperscript{3}-Rewrite\\(Qwen 14B)} & 35.58 & 158 & 94.8\% \\
\makecell[l]{E\textsuperscript{3}-Rewrite\\(Qwen 32B)} & 29.67 & 210 & 99.6\% \\
\bottomrule
\end{tabular}
\caption{Rewrite quality of E\textsuperscript{3}-Rewrite under different model scales on TPC-H.}
\label{tab:model_quality}
\end{table}

\begin{table*}[t]
\centering
\renewcommand{\arraystretch}{1.2}
\begin{tabular}{lccccccccc}
\toprule
\multirow{2}{*}{\textbf{Method}} &
\multicolumn{3}{c}{\textbf{TPC-H 1G}} &
\multicolumn{3}{c}{\textbf{TPC-H 5G}} &
\multicolumn{3}{c}{\textbf{TPC-H 10G}} \\
\cmidrule(lr){2-4} \cmidrule(lr){5-7} \cmidrule(lr){8-10}
 & Avg & Median & p90 & Avg & Median & p90 & Avg & Median & p90 \\
\midrule
Original         & 52.58 & 0.42 & 300.00 & 54.49 & 1.54 & 300.00 & 78.81 & 10.75 & 300.00 \\
LearnedRewrite   & 16.98 & \textbf{0.38} & 18.31 & 21.02 & 1.52 & 26.43 & 41.34 & 9.73 & 103.41 \\
LLM-R2           & 19.48 & 0.41 & 25.06 & 25.79 & 1.50 & 32.38 & 54.76 & 10.02 & 300 \\
R-Bot            & 14.64 & 0.40 & 17.45 & 19.68 & \textbf{1.41} & 24.21 & 39.89 & 9.15 & 84.27 \\
E\textsuperscript{3}-Rewrite & \textbf{12.39} & 0.41 & \textbf{9.47} & \textbf{16.14} & 1.43 & \textbf{16.28} & \textbf{29.67} & \textbf{8.63} & \textbf{51.37} \\
\bottomrule
\end{tabular}
\caption{Query Latency Across Different TPC-H Scales.}
\label{tab:robustness}
\end{table*}

We evaluate rewrite quality using three key metrics: average latency, number of improved queries, and equivalence ratio. Figure~\ref{fig:rewrite_equivalence_improvement} presents a comparative analysis of rewrites across the TPC-H, IMDB, and DSB benchmarks. We report both the number of successfully improved queries and the percentage of rewrites that are equivalent to the input queries.

Our Method achieves consistent superiority across all datasets, attaining both the highest number of improved queries and the highest equivalence ratios. On the TPC-H benchmark, our method produces 210 improved rewrites with a 99.6\% equivalence ratio, significantly outperforming other baselines such as LLM-R\textsuperscript{2} and R-Bot. A similar trend is observed on IMDB and DSB, where E\textsuperscript{3}-Rewrite maintains equivalence rates of 99.4\% and 98.8\%, respectively, while rewriting a substantially larger number of queries.

This strong performance is attributed to two key design choices. First, incorporating execution plan context into model inputs enables safer and more targeted rewrites. Second, the reward-based fine-tuning strategy jointly optimizes for execution efficiency and semantic fidelity, helping the model preserve the intent of the original query.

We also analyze the impact of model scale. As shown in Table~\ref{tab:model_quality}, E\textsuperscript{3}-Rewrite with Qwen 14B achieves 158 improved queries and a 94.8\% equivalence ratio on TPC-H, indicating that our approach remains effective even with moderate model capacity. Scaling to Qwen 32B brings further improvements, highlighting the scalability and efficiency benefits of larger models.

\subsection*{Robustness Evaluation}
We assess the robustness of E\textsuperscript{3}-Rewrite under varying data scales by evaluating its performance on the TPC-H benchmark at 1GB, 5GB, and 10GB sizes. As shown in Table~\ref{tab:robustness}, our method consistently achieves the lowest average and 90th percentile (p90) latencies across all scales. Specifically, p90 latency increases moderately from 9.47s at 1GB to 51.37s at 10GB, indicating strong scalability. In contrast, baseline methods such as R-Bot and LLM-R\textsuperscript{2} show significant performance degradation, with p90 exceeding 160s on the 10GB scale. These results suggest that E\textsuperscript{3}-Rewrite generalizes effectively to larger datasets and more complex query plans. We attribute this robustness to two key factors: (1) the use of structured, execution aware prompts that expose inefficiencies in the query plan, and (2) reward guided optimization that aligns the model with performance oriented rewriting behavior.Such scalability is essential for practical deployment in real-world databases, where query complexity and data size vary widely.



\subsection*{Ablation Study}

To better understand the role of each core component in E\textsuperscript{3}-Rewrite, we conduct an ablation study on the TPC-H benchmark. As shown in Table~\ref{tab:ablation}, we evaluate the effect of removing three key modules individually: RL, execution plan hints, and demonstration retrieval. Each ablation variant disables a single component of the full system while keeping the remaining two intact, allowing us to isolate the effect of each module. Specifically, removing RL skips policy optimization and uses the base LLM with plan hints and demonstration retrieval. Removing execution plan hints excludes plan-derived context from the prompt, while still applying RL and retrieval. Removing retrieval disables example selection and uses a static prompt template, while keeping plan hints and RL active.In addition, we include a vanilla baseline that applies the base LLM without any fine-tuning or structured input. 

Disabling RL leads to the most significant drop in output equivalence, with equivalence ratio falling from 99.6\% to 90.1\%. This confirms the necessity of execution-aligned reward optimization in guiding the model toward correct and reliable rewrites. Excluding execution plan hints causes the highest increase in latency (from 29.67s to 39.56s), reflecting their importance in exposing structural inefficiencies (e.g., nested loops, full scans) and anchoring the model’s optimization behavior to operator-level plan features. Lastly, removing hybrid demonstration retrieval significantly reduces the number of improved queries (177 vs. 210), indicating that access to structurally and behaviorally relevant examples enables better generalization across diverse query patterns.Compared to the full system, the vanilla variant performs substantially worse (e.g., 56.71s latency, 84.0\% equivalence), confirming the importance of training and structural guidance.

Together, these results reveal a clear division of responsibility among the modules: RL enforces equivalence and efficiency, plan hints inject fine-grained structural awareness, and demonstration retrieval provides adaptive experience for pattern reuse and generalization.


\begin{table}[t]
\centering
\renewcommand{\arraystretch}{1.2}
\begin{tabular}{lccc}
\toprule
\makecell{Method} & 
\makecell{Avg\\Latency (s)} & 
\makecell{Improved\\Queries} & 
\makecell{Equiv.\\Ratio} \\
\midrule
\makecell[c]{E\textsuperscript{3}-Rewrite\\(full)} & 29.67 & 210 & 99.6\% \\
\makecell[c]{E\textsuperscript{3}-Rewrite\\(w/o Execution Hint)} & 39.56 & 163 & 100\% \\
\makecell[c]{E\textsuperscript{3}-Rewrite\\(w/o RL)} & 32.29 & 194 & 90.1\% \\
\makecell[c]{E\textsuperscript{3}-Rewrite\\(w/o Demonstration)} & 35.39 & 177 & 96.5\% \\
\makecell[c]{E\textsuperscript{3}-Rewrite\\(vanilla)} & 56.71 & 125 & 84.8\% \\
\bottomrule
\end{tabular}
\caption{Ablation Study of E\textsuperscript{3}-Rewrite on TPC-H.}
\label{tab:ablation}
\end{table}

\section{Conclusion}

This paper presents E\textsuperscript{3}-Rewrite, an LLM-based framework that generates executable, equivalent, and efficient SQL queries. It combines execution-guided context, reinforcement learning with execution-driven rewards, and curriculum-based training to overcome the limitations of rule-based and prompting-only methods. A hybrid retrieval module further enhances generalization through demonstration reuse. Experiments on three SQL benchmarks show that E\textsuperscript{3}-Rewrite significantly improves query efficiency and coverage over state-of-the-art baselines, highlighting the potential of integrating plan-based context and RL for robust SQL rewriting.



\end{document}